\documentclass[12pt]{iopart}
\usepackage{hyperref}
\usepackage{graphicx}
\usepackage{epsfig}
\usepackage{dsfont}
\usepackage{verbatim}   
\usepackage[utf8]{inputenc}

\newcommand{\ket}[1]{\mathop{\left| #1 \right\rangle}\nolimits}

\begin{document}
 \title{QPSK coherent state discrimination via a hybrid receiver}

\author{C~R~M\"{u}ller$^{1,2}$, M~A~Usuga$^{3,1}$, C~Wittmann$^{1,2}$, M~Takeoka$^4$, Ch~Marquardt$^{1,2}$, U~L~Andersen$^{3,1}$ and G Leuchs$^{1,2}$ }
\address{	$^1$ Max Planck Institute for the Science of Light, Erlangen, Germany,\\
$^2$ Department of Physics, University of Erlangen-Nuremberg, Germany, \\
$^3$ Department of Physics, Technical University of Denmark, Kongens Lyngby, Denmark\\
$^4$ National Institute of Information and Communications Technology, 4-2-1 Nukui-Kita Koganei, Tokyo 184-8795, Japan}
\eads{\mailto{gerd.leuchs@mpl.mpg.de}, \mailto{ulrik.andersen@fysik.dtu.dk}}

\date{\today}

\begin{abstract}
We propose and experimentally demonstrate a near-optimal discrimination scheme for the quadrature phase shift keying protocol (QPSK). We show in theory that the performance of our hybrid scheme is superior to the standard scheme - heterodyne detection - for all signal amplitudes and underpin the predictions with our experimental results. Furthermore, our scheme provides the hitherto best performance in the domain of highly attenuated signals. The discrimination is composed of a quadrature measurement, a conditional displacement and a threshold detector.  
\end{abstract}


\maketitle

\section{Introduction}

It is one of the innermost consequences of the laws of quantum mechanics that non-orthogonal states can not be discriminated with certainty \cite{Helstrom67}. This allows for applications such as quantum key distribution (QKD) \cite{Bennett}, but it also ultimately limits the capacity in communication channels \cite{giovanetti}. 

In an optical communication protocol, a sender encodes information into one or more parameters of the light field. Such a parameter could for instance be the light's frequency, the phase or the amplitude. The prepared signal states are subsequently sent through an optical channel and directed to the receiver where the information is retrieved via an adequate measurement. However, if the power of the received signals is small, i.e. on the order of single photons, quantum mechanics has to be taken into account. In this regime, the minimum error rate for the discrimination of the signals is not only limited by the shortcomings of the technical apparatus but also by the laws of quantum mechanics. These laws impose strict bounds, depending on the implemented type of encoding, which cannot be overcome by any measurement device. 

A lot of attention has already been devoted to the development \cite{Tsujino10, Tsujino11} and characterization \cite{Paris} of optimal and near-optimal discrimination strategies for the elementary binary encoding into optical coherent states of the light field $\{\ket{\alpha},\ket{-\alpha}\}$, which allows to transmit one bit of information per state.
A more efficient encoding is provided by quadrature phase shift keying (QPSK), a technique which is widely used in wireless networks for mobile phones \cite{UMTS} and backbone fiber networks. The QPSK alphabet comprises four states equally separated by a phase of $\pi/2$ and  allows for the transmission of two bits of information per signal state $\{\ket{\alpha}, \ket{i \alpha}, \ket{-\alpha},\ket{-i \alpha}\} \Rightarrow \{00, 01,11, 10\}$. The minimal error rates for the discrimination of the QPSK alphabet have been derived by Helstrom \cite{Helstrom76, Osaki}. 

In the case of binary alphabets, it has been shown that the feasible secret key rates of quantum key distribution systems \cite{Bennett} can be largely improved by optimizing the receiver scheme \cite{Wittmann2010, Wittmann2010b}. Since QKD protocols with alphabets of four or higher number of states are also investigated \cite{Lorenz, Sych, Leverrier}, optimized receivers for such alphabets are of great interest.


In this paper, we present a novel discrimination scheme. We use a hybrid approach which means that we consider both fundamental representations of our quantum states: the discrete and the continuous representation. We prove in theory and provide experimental evidence that the standard scheme - heterodyne detection - can be outperformed for any signal amplitude.

Let us discuss different discrimination strategies for the QPSK alphabet. Besides heterodyne detection, where the received state is inferred from the beat signal between the signal and a local oscillator of slightly different frequency, there are two other advanced discrimination schemes, based on a photon counting detector and feedback that were proposed by Bondurant \cite{Bondurant}. In all these receivers, the measurement is performed by a single detection stage. In contrast, it is also possible to divide the state into parts which can be distributed among serveral detection stages. 
This method is for instance utilized in dual homodyne detection, where the received state is inferred by first splitting it on a balanced beam splitter and subsequently measuring the projections along two orthogonal quadratures via two homodyne detectors. However, the retrieved information in a dual homodyne measurement and in a heterodyne detection is identical such that the error rate is not reduced by the additional detection stage. It is for this equivalence that the terms heterodyne detection and dual homodyne detection are commonly used in a synonymic way. Recently, another receiver capable of achieving error rates below the heterodyne limit was proposed by Becerra et al. \cite{Migdall}. This scheme is based on successive measurements on parts of the state and feed forward.

\begin{figure}
\centering
\includegraphics[width=9cm]{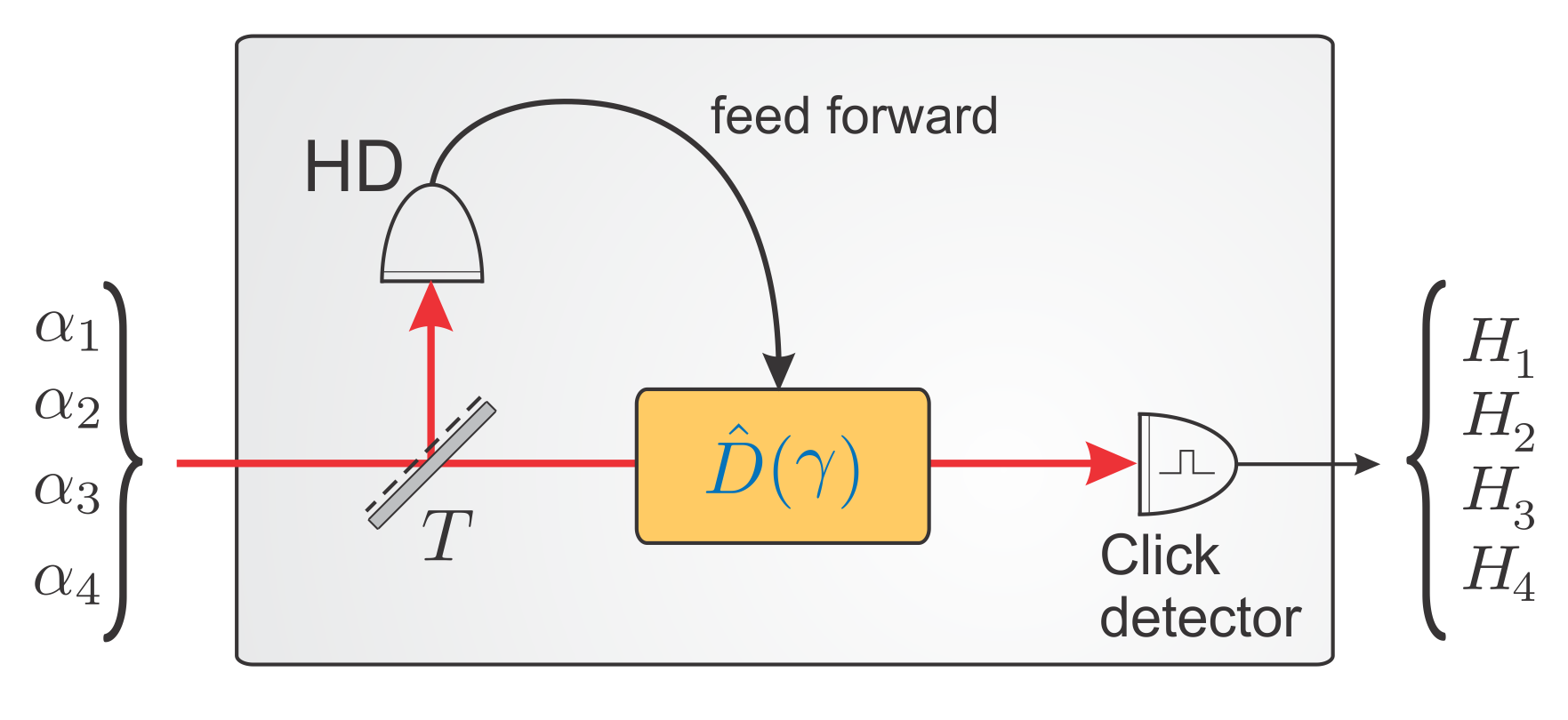}
\caption{(color online) Schematic of the hybrid discrimination scheme. First a homodyne detector distinguishes between pairs of states. The result is forwarded to a click detector stage, which is tuned for the discrimination of the remaining binary state.}
\label{DiscriminationScheme3}
\end{figure}

Our strategy is to perform two successive measurements on parts of the quantum state. The result of the first measurement reveals partial information about the state and is used to optimally tune the receiver for the second measurement. A schematic of the discrimination procedure is presented in Fig.\ref{DiscriminationScheme3}. The first measurement is performed by a homodyne detector (HD), best described by continuous variables. The homodyne measurement under a proper quadrature allows us to discard half of the possible states by making a binary decision based on the quadrature projections of the signal. The homodyne result is forwarded to a photon counting receiver, which finally identifies the input state by discriminating between the two remaining states. This task is performed near-optimally by an optimized displacement receiver \cite{Wittmann2008,TakeokaSasaki}, which is an advancement over the Kennedy receiver \cite{Kennedy}.


We implemented the hybrid scheme employing both the Kennedy (K) receiver and the optimized displacement (OD) receiver. The homodyne-Kennedy receiver (HD-K) beats the heterodyne detection for signal powers above a threshold (around $|\alpha|^2 \approx 1.6$). However the homodyne-optimized displacement receiver (HD-OD) outperforms heterodyne detection for any signal power.

\section{Description of the protocol}

Suppose, we are given a quadrature phase-shift keyed (QPSK) coherent signal $\ket{\alpha_n}, n\in\{ 1, 2, 3, 4\}$, where 
\begin{equation}\label{QPSKstatedefinition}
\alpha_n = |\alpha| \,e^{\,i\cdot (n-\frac{1}{2}) \cdot \frac{\pi}{2}}.
\end{equation}
and each of the states in the mixture has an \textit{a priori} probability of $p=1/4$.
The quantum limit -  the Helstrom bound \cite{Helstrom76}- for the discrimination of these signals is asymptotically given by $P_{err}^H = \frac{1}{2}\,e^{-2|\alpha|^2}$ for $|\alpha|^2\gg1$. 

The input signal is divided by a beam splitter (BS) with transmittance $T = t^2$ and reflectivity $R = r^2 = 1-t^2$. The transmitted and reflected parts are guided to the homodynde detector and the photon counting stage. First, one performs a homodyne detection along the P quadrature in phase space and makes a decision whether the signal is in the upper or the lower half plane. The result is forwarded to the photon counting receiver, which is then tuned for the discrimination of the remaining pair of states.

Let us recall the expression for the error probability in hypothesis testing:
\begin{equation}
P_{err} =\sum _{m\neq l} P(H_m|H_l) P(H_l) \nonumber
\end{equation}
In the case of QPSK $m,l\in \{1,2,3,4\}$ and the expression contains 12 terms corresponding to detection errors expressed by the conditional probabilities $P(H_m|H_l)$ which correspond to choosing the hypothesis $H_m$:``state $m$ was sent" when the correct hypothesis is $H_l$:``state $l$ was sent". 
In communications, the bit error rate (BER) is of particular interest. It is defined as the ratio between the number of erroneous bits and the total number of sent bits. For the QPSK alphabet the BER can explicitly be written as
\begin{equation}
\mathrm{BER}^{QPSK} = \frac{1}{4} \sum _{m\neq l} r_{m,l} P(H_m|H_l), 
\end{equation}
where $r_{m,l}=1$ for $|m-l|=2$ and $r_{m,l}=1/2$ otherwise. This means that higher bit errors are assigned to errors between distant states which will occur less frequently. However, in this work we will concentrate on the minimum error rate. 
In order to calculate the error probability $P_{err}$, it will be more convenient to first evaluate the success probabilities for the homodyne detector HD, the Kennedy receiver K and the optimized displacement receiver OD separately $P_{succ}^{HD,K,OD \, m}=P(H_m|H_m)$ and then simply find: $P_{err}^{HD-K} =1-P_{succ}^{HD-K}$ and $P_{err}^{HD-OD} =1-P_{succ}^{HD-OD}$, which have only 4 terms. In the case of the hybrid detectors analyzed here, the probability of success of the individual binary receivers is independent, and we can write:
\begin{eqnarray}
P^{\mathrm{HD}-\mathrm{K}}_{err}  &=&1-\sum _{m}p_mP^\mathrm{HD\,m}_{succ}P^\mathrm{K\,m}_{succ}   \nonumber \\
P^{\mathrm{HD}-\mathrm{OD}}_{err} &=&1-\sum _{m}p_mP^\mathrm{HD\,m}_{succ}P^\mathrm{OD\,m}_{succ} \nonumber 
\end{eqnarray}
where $p_m=P(H_m)=1/4$ are the a priori probabilities.

\begin{figure}
\includegraphics[width=16cm]{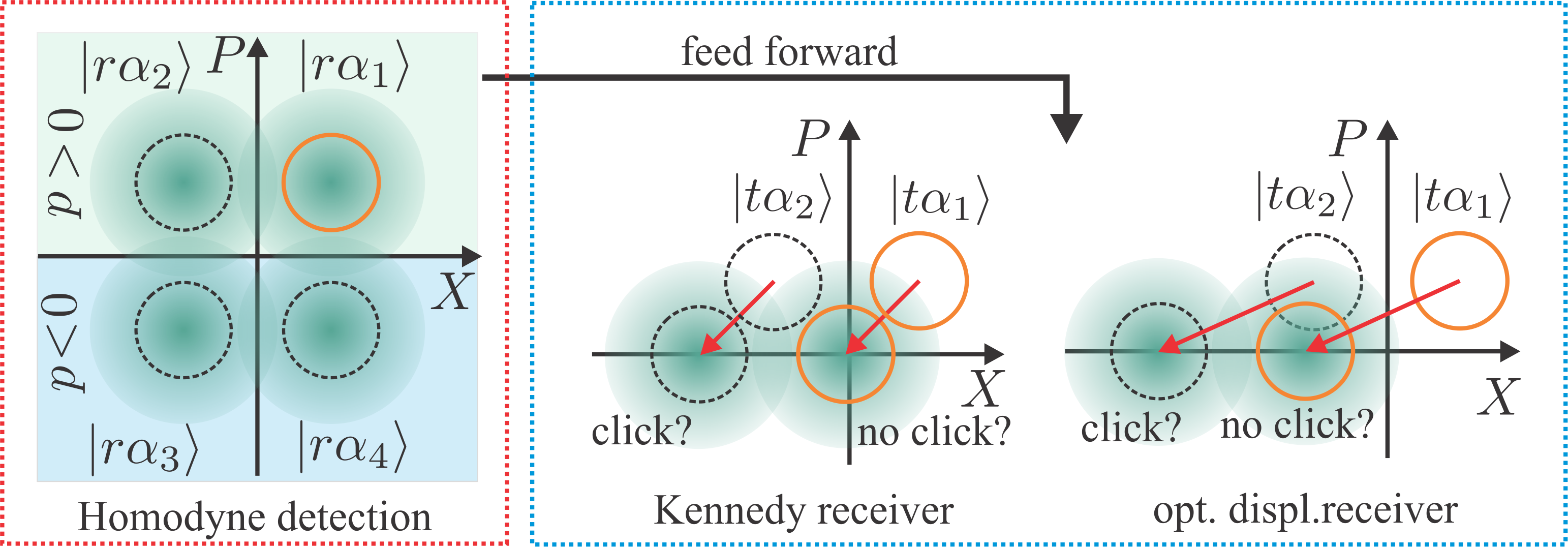}
\caption{(color online) Illustration of the measurements in phase space. The HD projects the states onto the P quadrature and forwards the measurement outcome to the click detector stage. Based on the forwarded information, the displacement prior to the click detector is tuned for the discrimination of the remaining pair of states. }
\label{DiscriminationScheme}
\end{figure}

Let us illustrate the procedure in more detail by assuming the signal is prepared in the state $\ket{\alpha_1}$ as indicated in Fig.\ref{DiscriminationScheme}. 
The reflected part $\ket{r\cdot\alpha_1}$ is directed to the homodyne detector, which discriminates between positive and negative values of the projection onto the P quadrature and is described by the POVM elements 
\begin{eqnarray}
\hat{\Pi}_{+}^{HD} &=&\int_{0}^{\infty}\mathrm{d}p | p\rangle \langle p| \nonumber \\ 
\hat{\Pi}_{-}^{HD} &=& 1- \hat{\Pi}_{+}^{HD}. 
\end{eqnarray}
The probability to observe the erroneous outcome $p\le0$ is given by

\begin{equation}
P_{err}^{HD} = 1-\int_{0}^{\infty} \left| \left\langle p\, \right|\left.r\cdot \alpha_1 \right\rangle \right| ^2  \mathrm{d}p\\ = \frac{1}{2}\left( 1-\mathrm{erf}\left[\sqrt{2}\,\left(r\frac{|\alpha|}{\sqrt{2}}\right) \right]\right).
\end{equation}

Note, that due to the projection onto the P quadrature, the effective signal amplitudes in the homodyne detection are reduced by a factor of $1/\sqrt{2}$. Supposing the measurement yielded the correct hypothesis, the next task is to discriminate between $|t\cdot \alpha_1 \rangle$ and  $|t\cdot \alpha_2 \rangle$ via the Kennedy or the optimized displacement receiver. 

For simplicity, let us first consider the Kennedy receiver. The signal is displaced such that one of the remaining candidate states is shifted to the vacuum state $|0\rangle$, while the other state gets amplified to an amplitude of $|\sqrt{2}\,t\cdot\alpha|$. The states are identified by observing whether or not a click occurs in the detector. In the scenario depicted in Fig.\ref{DiscriminationScheme}, the displacement was (arbitrarily) chosen to shift $|t\cdot\alpha_1\rangle$ to the vacuum. Therefore, the hypothesis is $|\alpha_1\rangle$, whenever no click was detected, whereas the input state is identified as $|\alpha_2\rangle$ if a detection event is recognized.
The corresponding POVM elements of the Kennedy receiver are 
\begin{eqnarray}
\hat{\Pi}_{\mathrm{no\, click}}^{K} &=& \hat{D}(t\cdot\alpha_1)|0\rangle \langle0| \hat{D}^{\dagger}(t\cdot\alpha_1) = |t\cdot\alpha_1\rangle \langle t \cdot\alpha_1| \nonumber \\
\hat{\Pi}_{\mathrm{click}}^{K} &=& \hat{\mathds{1}}-\hat{\Pi}_{\mathrm{no\, click}}^{K},
\end{eqnarray} 
where $\hat{D}(\cdot)$ denotes the displacement operator.
As the vacuum state is an eigenstate of the photon number operator ($\hat{n} = \hat{a}^{\dagger}\hat{a}$), the state shifted to the vacuum state will never generate a click and the error probability $P_{err}^{K\,1}= 1-\mathrm{Tr}[\hat{\Pi}_{\mathrm{no\, click}}^{K}|t\alpha_1 \rangle \langle t\alpha_1 |]$ is zero. The total error probability $P_{err}^1$ for correctly guessing $|\alpha_1 \rangle$ is then given by 

\begin{equation}
P_{err}^1 = 1- \left(1-P^{HD}_{err}\right) \cdot \left(1-P_{err}^{K\,1}\right) = P^{HD}_{err}.
\end{equation}

If instead the input signal was $\ket{\alpha_2}$ (or equivalently if the displacement was chosen to shift $\ket{t\cdot\alpha_2}$ to the vacuum state), the error probability of the Kennedy receiver is $P_{err}^{K\,2} = 1- \mathrm{Tr}[\hat{\Pi}_{\mathrm{click}}^{K}|t\alpha_2 \rangle \langle t\alpha_2 |] = e^{-2|t\,\alpha|^2}$, where the errors originate from the remaining overlap between the displaced state and the vacuum state. 
The total error rate for the detection of $\ket{\alpha_2}$ is given by

\begin{equation}
P_{err}^2 =  1 - \left( 1-P^{HD}_{err}\right) \left(1-P^{K\,2}_{err}\right)  = 1 - \left( 1-P^{HD}_{err}\right) \left( 1- e^{-2\,|t\,\alpha|^2} \right) , 
\end{equation}

Note, that the same error rates follow for the other signals (n = 3, 4). Consequently, the average error probability for the HD-K hybrid receiver is
\begin{eqnarray}\label{erroroverall}
P_{err}^{HD- K} & = & \frac{1}{2} \left( 
                1 - \left( 1 - P^{HD}_{err} \right)\cdot \left(1 - P^{K\,1}_{err} \right)  
              + 1 - \left( 1 - P^{HD}_{err} \right)\cdot \left(1 - P^{K\,2}_{err} \right) \right) \nonumber  \\
		      & = & 1 - \frac{1}{2} (1+\mathrm{erf}\left[|r\,\alpha|\right])( 1- \frac{1}{2} e^{-2|t\,\alpha|^2} ),
\end{eqnarray}

where $P_{err}^{K} = \frac{1}{2} e^{-2|t\,\alpha|^2}$ is the average error rate of the Kennedy receiver stage. 
\begin{figure}
\centering
\includegraphics[width=8cm]{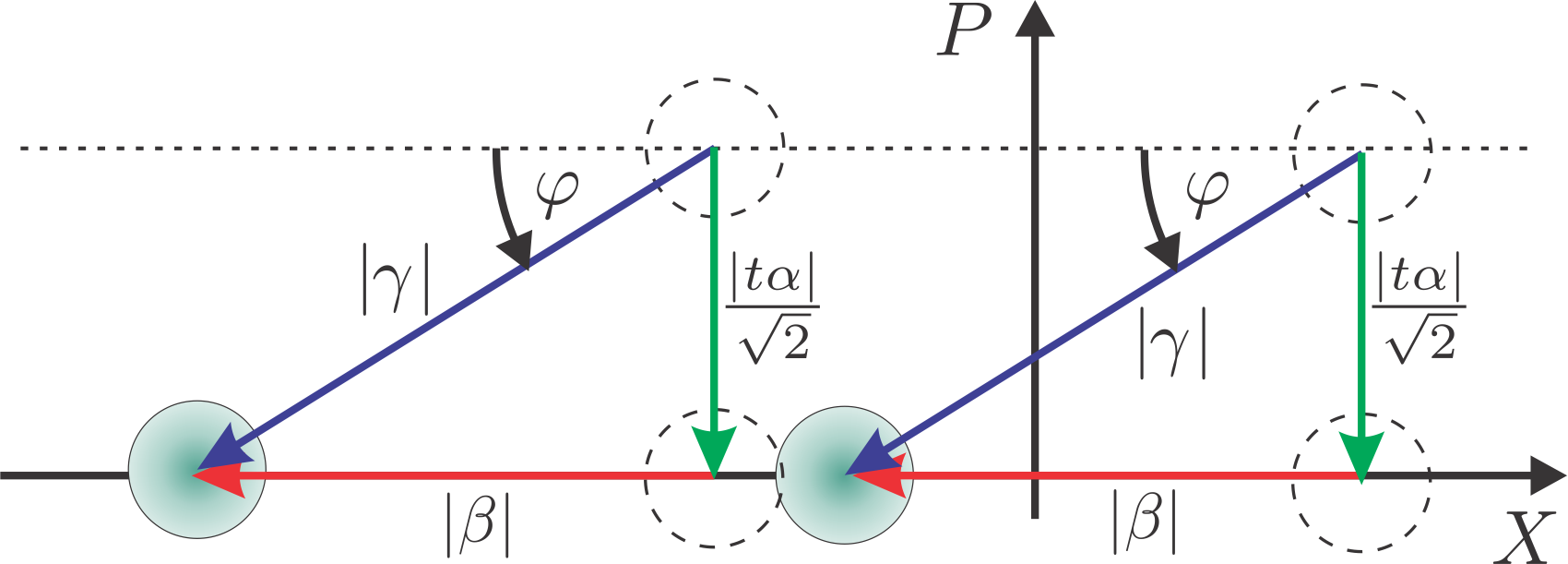}
\caption{(color online) Sketch illustrating the two elementary displacement operations from which the optimal displacement for the QPSK alphabet is derived.}
\label{displacementsketch}
\end{figure}

The error rates of the Kennedy receiver can however be lowered by optimizing the displacement, which leads to the optimized displacement (OD) receiver. The error rates of the Kennedy- and the optimized displacement receiver for the discrimination of binary states have been compared to the optimal Gaussian approach (homodyne detection) in \cite{Wittmann2008}. 
The Kennedy receiver is superior to homodyne detection for signals with a mean photon number $\bar{n} > 0.4$, whereas the optimized displacement receiver outperforms the optimal Gaussian approach for any signal power. 

To derive the optimal displacment parameter $\gamma$ for the QPSK signal it is convenient to separate the total displacement into two elementary steps as illustrated in Fig.\ref{displacementsketch}. First, the states are displaced to the X quadrature, which is described by the displacement operator $\hat{D}(-i\,|t\,\alpha|/\sqrt{2})$. The situation is then equivalent to a binary state discrimination problem for two states with amplitude $|\alpha|/\sqrt{2}$. In this configuration, the optimal displacement $\beta$ is given by the solution of the transcendental equation \cite{TakeokaSasaki}
\begin{equation}
\frac{t}{\sqrt{2}} \alpha = \beta \tanh\left(\sqrt{2}\,t\alpha\beta\right),
\end{equation}
which is obtained by requiring $\partial P_{err}^{OD}/ \partial \beta = 0$. 

As illustrated in Fig.\ref{displacementsketch}, the optimal displacement amplitude $|\gamma|$ and phase $\varphi$ for the QPSK signal are then following as
\begin{eqnarray}
|\gamma | & = & \sqrt{\frac{|t\,\alpha|^2}{2} + |\beta|^2 }  \\
\varphi  & = &  \mathrm{atan}\left({\frac{|t\,\alpha|}{\sqrt{2}|\beta|}}\right). 
\label{displparams}
\end{eqnarray}

Combining the two elementary displacements, the OD receiver is finally described by the POVMs 
\begin{eqnarray}
\hat{\Pi}_A^{OD} & = &  
\hat{D}\left(i\,|t\,\alpha|/\sqrt{2}\right)\hat{D}(\beta)|0\rangle\langle0|\hat{D}^{\dagger}(\beta)\hat{D}^{\dagger}\left(i\,|t\,\alpha|/\sqrt{2}\right) \nonumber \\
     & = &  
\hat{D}(\gamma)|0\rangle \langle0| \hat{D}(\gamma)^{\dagger} = |\gamma\rangle \langle\gamma| 
\end{eqnarray}
and $\hat{\Pi}_B^{OD} = \hat{\mathds{1}}-\hat{\Pi}_A^{OD}$, with $\gamma=\beta+i\,|t\,\alpha|/\sqrt{2}$.

The error rates for the HD-OD hybrid receiver follow directly by exchanging the Kennedy error rates $P_{err}^{K\,1,2}$ for the error rates of the OD receiver $P_{err}^{OD\,1,2}$. The total error rate is then given by  
\begin{eqnarray}
P^{OD}_{err} &=& \frac{1}{2}\left( \mathrm{Tr}\left[\hat{\Pi}_{\mathrm{A}}^{OD}\,|t\alpha_1 \rangle \langle t\alpha_1 |\right] + \mathrm{Tr}\left[\hat{\Pi}_{\mathrm{B}}^{OD}\,|t\alpha_2 \rangle \langle t\alpha_2 |\right]  \right) \nonumber \\
             &=& \frac{1}{2} - \exp\left(-t^2\frac{|\alpha|^2}{2} + |\beta|^2 \right) \sinh\left({\sqrt{2}\,t\,\alpha\beta} \right).
\end{eqnarray}

\begin{figure}[htb]
\includegraphics[width=16cm]{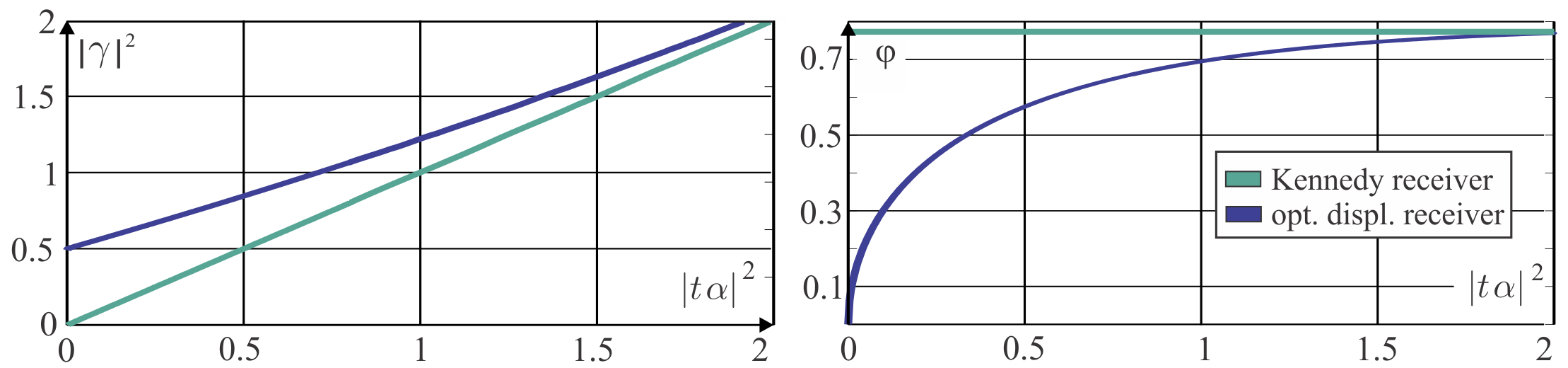}
\caption{(color online) Optimal absolute values for the displacement $|\gamma|^2$ and optimal displacement phases $\varphi$ in dependence of the transmitted part of the signal $|t\,\alpha|^2$ for both the Kennedy and the optimized displacement receiver. }
\label{opt_beta}
\end{figure}

The optimal displacement parameters for the Kennedy receiver and the OD receiver are shown as a function of the transmitted signal in Fig.\ref{opt_beta}. The displacement in the OD receiver is clearly increased for small signal powers and has a minimum value of $|\gamma|^2=0.5$ in the limit of very low signal powers. Asymptotically, the displacement of the OD receiver approaches the values of the Kennedy receiver, which is identical to the transmitted signal power. 
The phase $\varphi$ describes the direction of the  displacement in phase space as sketched in Fig.\ref{displacementsketch}. In case of bright signals, both detectors displace the states towards the vacuum state, which corresponds to a phase of $\varphi = \pi/4$. With decreasing signal power the phase in the OD receiver is asymptotically approaching $\varphi=0$, which corresponds to a displacement parallel to the X quadrature.

\begin{figure}[htb]
\centering
\includegraphics[width=8cm]{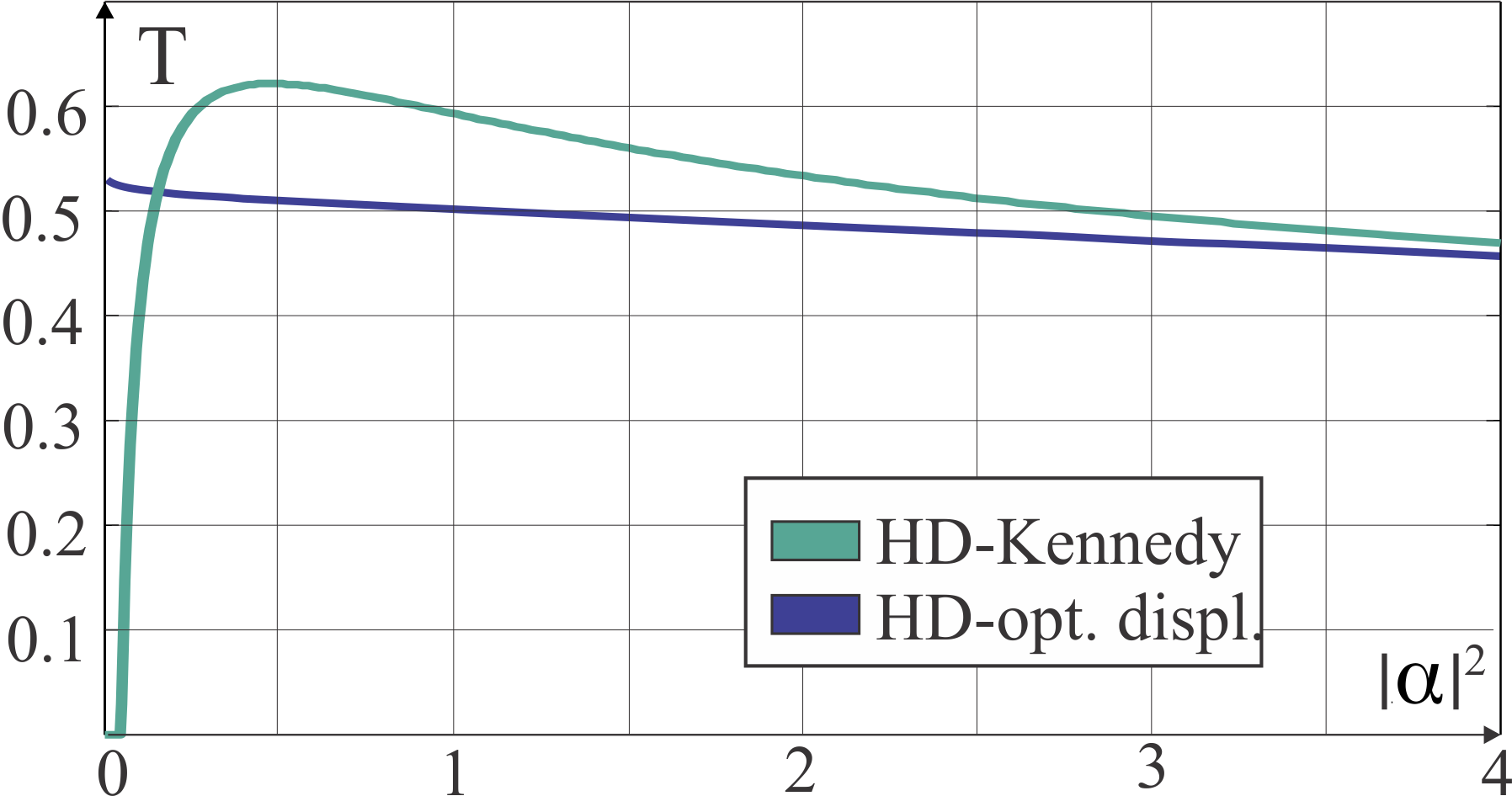}
\caption{(color online) Optimal parameters for the transmittance to the photon detection stage in case of the Kennedy- and the optimized displacement receiver.}
\label{transmission}
\end{figure}

Besides of the displacement parameter $\gamma$, the transmittance $t^2$ of the beam splitter can be optimized to minimize the error rate. The optimal parameters are shown in Fig.\ref{transmission}. In case of small signal powers  $|\alpha|^2\approx1$, the quantum state in the HD-OD receiver is distributed nearly equally among the two receiver stages  $t^2\approx0.5$. With increasing signal power the share of the photon counting receiver is monotonically decreasing. In contrast, the optimized transmission for the HD-K receiver shows a distinct maximum around $|\alpha|^2\approx0.5$, but approaches the optimal transmission parameter of the HD-OD receiver asymptotically with increasing signal power. In the limit of very high signal powers $|\alpha|^2\gg1$ (not shown in the figure), the share of the photon counting receivers tends to $t^2=0$. This reflects the increasing imbalance between the performance in binary state discrimination of the photon counting receivers compared to HD detection \cite{Wittmann2008}. In this regime, the photon counting receivers' performance is (in theory) exceedingly superior to the quadrature measurements. The homodyne detection thus constitutes the main source of errors. The total error is minimized by allocating the major share of the state to the HD detector. Practically however, the performance of click detectors in the high amplitude regime is technically limited by dark counts. 

\section{Experimental setup}

\begin{figure}
\centering
\includegraphics[width=8cm]{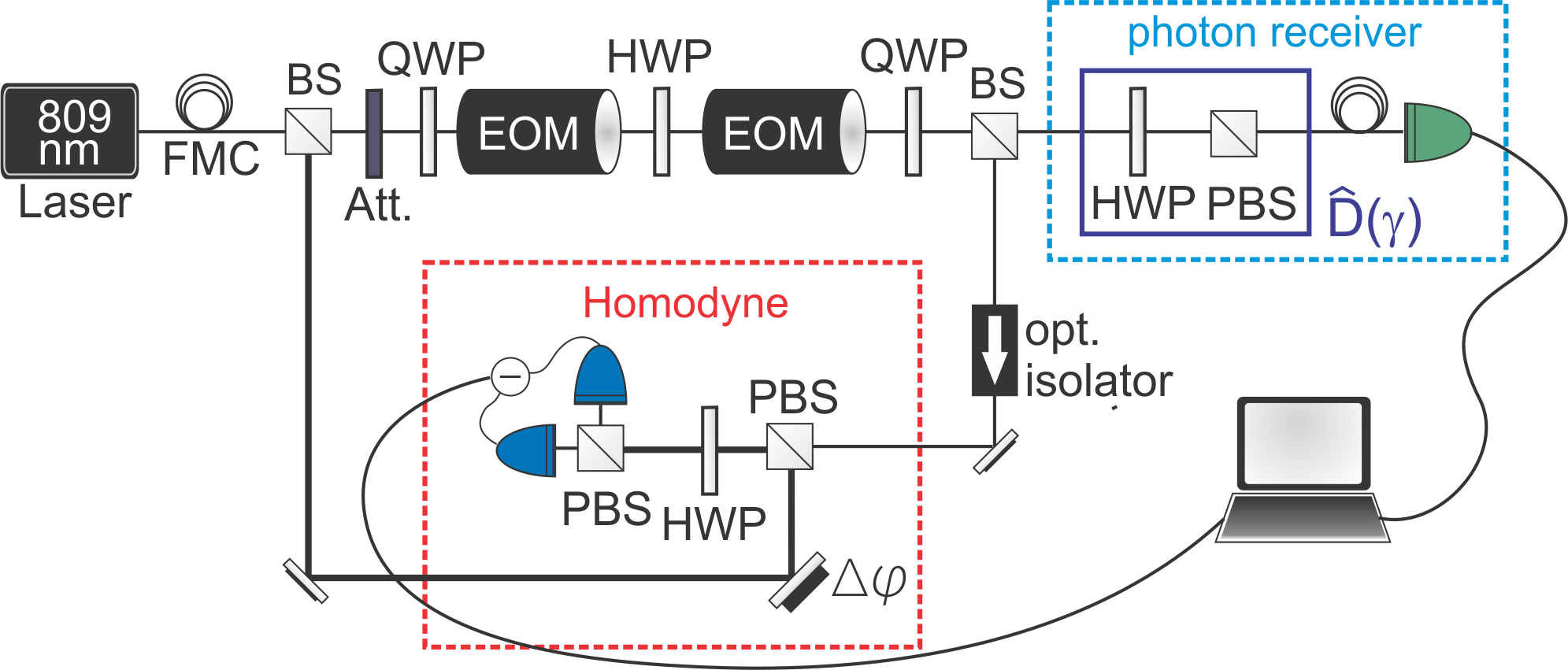}
\caption{(color online) Experimental setup for the discrimination of the QPSK coherent states.}
\label{setup}
\end{figure}

We proceed with a description of the experimental setup which is shown in Fig.~\ref{setup}. Our source is a grating-stabilized diode laser operating at a wavelength of $809\,$nm. The laser has a coherence time of $1\,\mu$s and is measured to be shot noise limited within the detection bandwidth. First, the beam passes a single mode fiber to purify the spatial mode profile. Subsequently, the beam is split asymmetrically into two parts: a bright local oscillator (LO), which is directed to the HD stage and a weak auxiliary oscillator (AO), which is used both to prepare the signal states and to realize the displacement at the photon counting receiver stage. Directly after the first beam splitter, the AO passes an attenuator (Att.) to reduce its intensity to the few photon level. The use of of electro-optical modulators (EOMs) and wave plates allows to generate signal states as pulses of $800\,$ns and at a repetition rate of $100\,$kHz in the same spatial mode as the AO but with an orthogonal polarization.

The signal is split on a beam splitter 
and the parts are guided to the homodyne detector and the photon counting receiver, respectively. In the HD path, the signal mode is separated from the AO via an optical isolator aligned to absorb the remaining AO. Moreover, the isolator avoids back-propagation of photons from the LO to the photon counting receiver. Subsequently, the signal is spatially superposed with the LO on a polarizing beam splitter (PBS). Up to this point signal and LO are still residing in orthogonal polarization modes. The required interference is achieved by a combination of a half-wave plate HWP and a PBS. The wave plate is aligned to rotate the polarization axis by an angle of $\pi/4$. At this point, the signal and the LO have equal support on the principal axis of the subsequent PBS, such that they are split symmetrically and the interference is achieved. The measured quadrature in the HD is adjusted via a feed back controlled piezo-electric transducer in the LO path. The measured visibility between the signal and the LO is $V = 95\%$ and the quantum efficiency of the photo diodes is measured to be $ \eta_{\mathrm{diodes}} = 92 \pm 3 \% $. From this, the total quantum efficiency of the homodyne detection follows as $ \eta_{HD} = V^2\cdot\eta_{\mathrm{diodes}} = 83\pm3\%$.

In the photon counting receiver path the displacement is generated by coupling photons from the AO to the orthogonally polarized signal mode. This is achieved by first rotating the polarization of the signal and the AO via a HWP, followed by a projection onto the original signal polarization mode by a PBS. The angle of the HWP, and hence the displacement strength, is controlled by a stepper motor. If the required rotation angle $\theta$ is small, i.e. for a sufficiently bright AO, the disturbance of the signal states is small and the operation is equivalent to a perfect displacement operation. The displacement operation can be described as $\ket{\alpha} \ket{AO} \stackrel{HWP}{\rightarrow} \ket{\cos(\theta) \alpha + \sin(\theta)AO} \ket{AO'}  \stackrel{\cos(\theta)\approx 1}{\rightarrow}   \ket{\alpha+\gamma} \ket{AO'}$, where $\ket{AO}$ denotes the coherent state in the auxiliary oscillator mode. Experimentally however, increasing the AO power results in an increased dark count rate originating from the limited extinction ratio of the EOMs, which is measured to be $C \approx 1/500$. We therefore adjusted the mean photon number in the AO to optimize the trade off between state disturbance and dark count rate, which leads to an AO with about 20 photons. Finally, the displaced signal is coupled to a multi-mode fiber connected to an avalanche photo diode (APD). The APD is operated in an actively gated mode and has a measured quantum efficiency of  $\eta_{APD} = 63 \pm 3 \%$.

We probe the receiver with a sequence of test signals. Each sequence is composed of an initial block of phase calibration pulses used to lock the quadrature in the homodyne measurement, followed by 9 blocks of probe pulses. Each block contains the full QPSK alphabet for 34 different amplitudes in the range $|\alpha|^2 \in [0, 1.8]$. The stepper motor controlling the displacement is actuated after every 4000 runs of the sequence to vary the displacement $|\gamma|^2$. The results of the individual measurements are sent to a computer and saved. The feed forward is emulated in the post-processing, where only the data in which the adjusted displacement concurred with the result of the HD measurement is evaluated. 
A limitation in performing the displacement by means of a HWP is that the direction of the displacement is restricted along one specific quadrature, depending on the relative phase $\phi$ of the AO with respect to the signal mode. However, in order to fulfill the optimality criterion in the HD-OD receiver (see Fig.\ref{displacementsketch}), the direction of the displacement has to be adjusted depending on the signal amplitude (see Eq.(\ref{displparams}) and Fig.\ref{opt_beta}). To account for this requirement, the signals in the probe blocks are generated with an equidistantly varying relative phase to the AO in the range $\phi \in  \left[0,\pi/4\right]$.

The aim of the experiment is to provide a proof-of-principle demonstration of the hybrid receivers' performance unaffected by any imperfections of the implemented hardware, but only limited by the physical concept. In the analysis of the experimental data, we therefore assume unit quantum efficiencies for the individual receivers. 
Losses and detection inefficiencies, which can also straightforwardly be described as loss, merely result in a linear rescaling of the states' amplitudes. By combining this with the linearity of a beam splitter interaction, we can assign the detection inefficiencies to the state generation stage. This trick has proven to ease the understanding of the protocol by removing unnecessary prefactors \cite{Geremia07}. The assignment leads to a beam splitter with an effective splitting ratio:
$T \rightarrow T' = \eta_{\mathrm{APD}}\,T/\left(\eta_{\mathrm{APD}}\,T + \eta_{\mathrm{HD}}\,R \right)$ and $R \rightarrow  R' = \eta_{\mathrm{HD}}\,R/\left(\eta_{\mathrm{APD}}\,T + \eta_{\mathrm{HD}}\,R \right)$.

\section{Experimental Results}

\begin{figure}[htb]
\includegraphics[width=16cm]{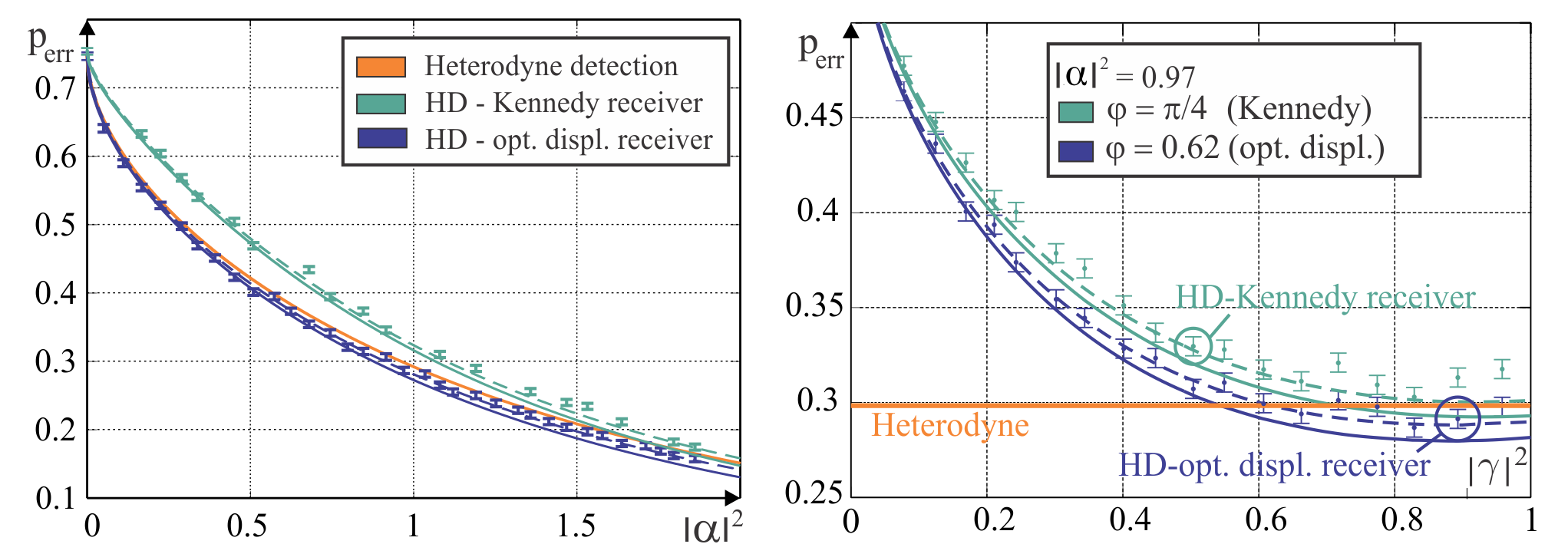}
\caption{(color online) (left) Experimental results for the error rates of the hybrid receivers compared to a perfect heterodyne detector. The dashed lines correspond to the theoretical prediction including the detrimental effects of dark counts.(right) Dependence of the hybrid receivers' error rates on the displacement amplitude for a input state with $|\alpha|^2=0.97$. The curves differ in the direction of the displacment in phase space.  }
\label{error_rate}
\end{figure}

We measured the error rates for both the HD-K receiver and the HD-OD receiver at an effective splitting ratio of T/R = 53/47. The results are compared to the performance of an ideal heterodyne receiver in Fig.\ref{error_rate}(left). The solid curves correspond to the theoretical error rates under ideal conditions, whereas the dashed curves include the detrimental effects of dark counts, which occurred with the probability of 2,72$\%$. The error bars were derived by error propagation of the experimental uncertainties of the input amplitude $\Delta|\alpha| = 0.01$ and the displacement amplitude $\Delta|\gamma| = 0.039$ as well as the fluctuations among repeated realizations of the experiment which were around 0.5\%. 

We find the error rates for the HD-K receiver by evaluating the data where the signal power of the displaced state  $|\alpha - \gamma|^2$ is minimal, i.e. when one state in the photon counting stage has been displaced to the vacuum. 
The error rates for the HD-OD receiver are derived by minimizing the error rate over the range of measured displacements $|\gamma|^2$ and displacement phases $\varphi$. 
The results for both receivers are in good agreement with the theoretical predictions. The measured error rates for the HD-OD receiver are below the corresponding error rate of the ideal heterodyne detector for any input amplitude. Moreover, most of the measurements beat the heterodyne receiver's performance with about one standard deviation.

The essential difference between the HD-K and the HD-OD receiver is illustrated in Fig.\ref{error_rate}(right), where the dependence of the error rates over the displacement is shown for an input signal with mean photon number $|\alpha|^2=0.97$. The curves differ in the respective displacement angles in the two receivers. While the HD-K receiver was measured at $\varphi = \pi/4$, the phase in the HD-OD receiver was adjusted to fulfill the optimality criterion (see Fig.\ref{displacementsketch}) corresponding to $\varphi = 0.62$. The configurations for the HD-K ($|\alpha|^2=|\gamma|^2$) and the HD-OD receiver (minimal error rate) are highlighted. Obviously, the performance of the HD-K receiver can already be enhanced by increasing the displacement amplitude $|\gamma|^2$, however the minimal error rates are only achieved if both the displacement amplitude and phase are optimized. The corresponding error rate for the standard heterodyne receiver is shown as a reference and is surpassed by the HD-OD receiver for a wide range of displacement amplitudes. The curvature of the error rate around the minimum is remarkably flat, such that the dependence on the absolute amplitude of the displacement $|\gamma|^2$ is low. 

\begin{figure}[htb]
\centering
\includegraphics[width=11cm]{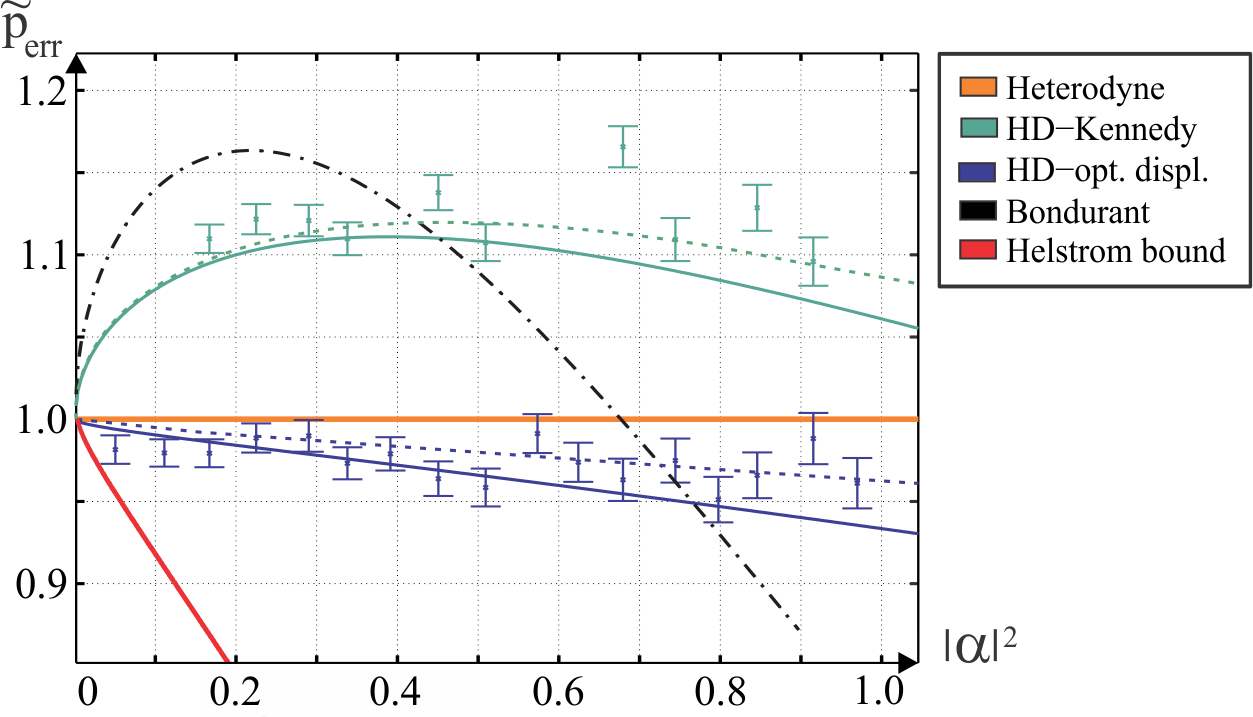}
\caption{(color online) Comparison of the error rates from different receivers normalized to the standard scheme - heterodyne detection. Solid lines correspond to the error rates under ideal conditions, while the dashed lines include the detrimental effects of dark counts. The Bondurant receiver is shown as a dashed-dotted curve and the quantum limit - the Helstrom bound - is illustrated by the red curve.}
\label{comparison}
\end{figure}

The relative error rates $\widetilde{p}_{err}$ of the hybrid receivers, normalized to the error rates of heterodyne detection are shown Fig.\ref{comparison}. Additionally, the relative error rates of the before mentioned Bondurant receiver \cite{Bondurant} is shown. Bondurant had proposed two similar discrimination schemes which he termed type I and type II, respectively. The curve shown in the figure correponds to the Bondurant reveiver of type I, which provides the better performance in the considered region. While this receiver outperforms heterodyne detection and also our hybrid approaches for conventional signal amplitudes, it can not provide an enhanced performance in the domain of highly attenuated signals. The HD-OD receiver provides to the best of our knowledge the hitherto minimal error rates for signals with mean photon numbers $|\alpha|^2 \leq 0.75$.

\section{Conclusion}
We have proposed and experimentally realized a hybrid quantum receiver for the discrimination of QPSK coherent signals. We showed experimentally, that our novel receiver can outperform the standard scheme - heterodyne detection - for any signal amplitude.

\section*{Acknowledgements}
This work was supported by the DFG project LE 408/19-2 and by the Danish Research Agency (project no. FNU 09-072632).

\section*{References}
\bibliographystyle{unsrt}

\end{document}